\documentclass[12pt]{article}

\usepackage{epsfig,graphicx}

\begin{document}

\centerline{\large Present Status of Our Knowledge of $|V_{cb}|$}
\vskip 1 cm
\def\vcb{|V_{cb}|}
\def\dec{\rightarrow}
\def\fvcb{${\cal F}(1)|V_{cb}|$}
\def\ups{$\Upsilon (4\rm S)$}

\centerline{M. Artuso\footnote{ email=artuso@physics.syr.edu;
  homepage=http://www.phy.syr.edu/~artuso.}}
  \centerline{\it Syracuse University, Syracuse, NY13244, USA}

\vskip 1 cm \centerline{\it Talk given at BEAUTY 2003, Pittsburgh,
PA, October 2003}

\begin{abstract}
The Cabibbo-Kobayashi-Maskawa parameter $|V_{cb}|$ plays an
important role among the experimental constraints of the Yukawa
sector of the Standard Model. The present status of our knowledge
will be summarized with particular emphasis to the interplay
between theoretical and experimental advances needed to improve
upon present uncertainties.
\end{abstract}


\bigskip

\section{Introduction}
In the framework of the Standard Model, the quark sector is
characterized by a rich pattern of flavor-changing transitions,
described by the Cabibbo-Kobayashi-Maskawa (CKM) matrix:
\begin{equation}
V_{CKM} =\left(\begin{array}{ccc}
V_{ud} &  V_{us} & V_{ub} \\
V_{cd} &  V_{cs} & V_{cb} \\
V_{td} &  V_{ts} & V_{tb} \end{array}\right).
\end{equation}
Since the CKM matrix must be unitary, it is determined by only
four independent parameters. Wolfenstein proposed an approximate
parameterization \cite{wolf-prl} that reflects the hierarchy
between the magnitude of matrix elements belonging to different
generations. Very frequently it is quoted in the approximation
valid only to $\lambda^3$. We need to carry out this expansion
further in order to incorporate  CP violation in neutral $K$
decays. This expression, accurate to $\lambda^3$ for the real part
and $\lambda^5$ for the imaginary part, is given by:

\begin{equation}
{\footnotesize \left(\begin{array}{ccc}
1-\lambda^2/2 &  \lambda & A\lambda^3(\rho-i\eta (1-\lambda^2/2)) \\
-\lambda &  1-\lambda^2/2-i\eta A^2\lambda^4 & A\lambda^2(1+i\eta\lambda^2) \\
A\lambda^3(1-\rho-i\eta) &  -A\lambda^2 & 1\end{array}\right).}
\end{equation}
The parameter $\lambda$ is well measured as 0.2196$\pm$0.0023
\cite{PDG2002}, constraints exist on $\rho$ and $\eta$ from
measurements of $V_{ub}$ and $B^0\bar{B}^0$ mixing. This paper
focuses on the magnitude of the CKM element  $|V_{cb}|$, related
to the Wolfenstein parameter $A$ \cite{wolf-prl}.

\section{$|V_{cb}|$ from the exclusive decay $B\rightarrow D^{\star} \ell \bar{\nu}$.}

HQET predicts that the differential partial decay width for this
process, $d\Gamma/dw$, is related to $\vcb$ through:
\begin{equation}
\frac{d\Gamma}{dw}(B\rightarrow D^{\star}\ell \nu) = \frac{G_F^2
|V_{cb}|^2}{48\pi^3}{\cal K}(w){\cal F}(w)^2,
\end{equation}
where $w$ is the inner product of the $B$ and $D^{\star}$ meson
4-velocities, ${\cal K}(w)$ is a known phase space factor and the
form factor ${\cal F}(w)$ is generally expressed as the product of
a normalization factor ${\cal F}(1)$ and $g(w)$, the Isgur-Wise
function, whose shape is constrained by dispersion relations
\cite{grinstein}. The analytical expression of $g(w)$ is not known
a-priori, and this introduces an additional uncertainty in the
determination of ${\cal F}(1)|V_{cb}|$. First measurements of $
|V_{cb}|$ were performed assuming a linear approximation for
${\cal F}(w)$. It has been shown\cite{Stone} that this assumption
is not justified, and that linear fits systematically
underestimate the extrapolation at zero recoil ($w=1$) by about
3\%. Most of this effect is related to the curvature of the form
factor, and does not depend strongly upon the details of the
non-linear shape chosen\cite{Stone}. All recent results use a
non-linear shape for ${g}(w)$, approximated with an expansion near
$w=1$\cite{CLN}, and is parameterized in terms of the variable
$\rho^2$, which is the slope of the form factor at zero recoil
given in \cite{CLN}.

Considerable theoretical work has been devoted to the parameter
${\cal F}(1)$. Ultimately a precise value for it may be determined
by lattice gauge calculations. Presently only a quenched lattice
evaluation is available and gives $0.913^{+0.024}_{-0.017}\pm
0.016^{+0.003}_{-0.014}~^{+0.000}_{-0.016} ~^{+0.006}_{-0.014}$.
The errors reflect the statistical accuracy, the matching error,
the finite lattice size, the uncertainty in the quark masses and
an estimate of the error induced by the quenched approximation,
respectively. The central value obtained with OPE sum rules is
similar, with an error of $\pm 0.04$ \cite{ckm-workshop}.
Consequently, I will use ${\cal F}(1)=\ 0.91 \pm 0.04$
\cite{ckm-workshop}.

\begin{table}
\begin{tabular}{lccc}
\hline {Experiment}
  & {${\cal F}(1)|V_{cb}|(\times 10^{3})$}
  & {$\rho^2$ }
  & {$\rm Corr_{stat}$}   \\ \hline
 ALEPH update    &  33.6$\pm$  2.1$\pm$ 1.6 & 0.75$\pm$ 0.25$\pm$ 0.37 & 94\% \\
 OPAL(partial reconstruction)&  38.4$\pm$  1.2$\pm$ 2.4 & 1.25$\pm$ 0.14$\pm$ 0.39 & 77\%\\
 OPAL (excl)          &  39.1$\pm$  1.6 $\pm$ 1.8 & 1.49$\pm$ 0.21$\pm$ 0.26& 95\% \\
 DELPHI (partial reco.)    &  36.8$\pm$  1.4$\pm$ 2.5 & 1.52$\pm$ 0.14$\pm$ 0.37 & 94\% \\
 DELPHI (excl. prelim.)    &  38.5$\pm$  1.8$\pm$ 2.1 & 1.32$\pm$ 0.15$\pm$ 0.34 & 89\% \\
 Belle           &  36.7$\pm$  1.9$\pm$ 1.9 & 1.45$\pm$ 0.16$\pm$ 0.20 & 91\% \\
 CLEO            &  43.6$\pm$  1.3$\pm$ 1.8 & 1.61$\pm$ 0.09$\pm$ 0.21 & 87\% \\
 BaBar           &  34.1$\pm$  0.2$\pm$ 1.3 & 1.23$\pm$ 0.02$\pm$ 0.28 & 92\% \\
\hline
World average&  36.7 $\pm$ 0.8 & 1.44 $\pm$ 0.14  & 91\%\\
\hline
\end{tabular}
\caption{Experimental results for \fvcb\ and $\rho ^2$ rescaled to
common inputs \cite{hfag}}
\end{table}

 The
main contributions to the ${\cal F} (1) \vcb$ systematic error in
the LEP results come from the uncertainty on the $ B\dec
D^{\star\star}\ell \nu$ shape and $ {\cal B}( b\to B_d)$,
($0.57\times 10^{-3}$), fully correlated among these experiments,
the branching fraction of $D$ and $D^{\star}$ decays, ($0.4\times
10^{-3}$), fully correlated among all the experiments, and the
slow pion reconstruction from Belle, CLEO, and BaBar which are
uncorrelated. The main contribution to the $\rho^2$ systematic
error is from the uncertainties in the measured values of R$_1$
and R$_2$ (0.12), fully correlated among all the experiments.
Because of the large contribution of this uncertainty to the
non-diagonal terms of the covariance matrix, the averaged $\rho^2$
is higher than one would naively expect.

Using ${\cal F} (1) = 0.91 \pm 0.04$ \cite{ckm-workshop}, this
method gives $\vcb = ( 40.2 \pm 0.9_{exp} \pm 1.8_{theo}) \times
10^{-3}$. The dominant error is theoretical, but there are good
prospects that lattice gauge calculations will significantly
improve their accuracy.

\section{$|V_{cb}|$ from the exclusive decay $B\rightarrow D \ell \bar{\nu}$.}

The study of the decay $B\dec D \ell \nu$ poses new challenges
both from the theoretical and experimental point of view.

The differential decay rate for $B\dec D \ell \nu$ can be
expressed as:
\begin{equation}
\frac{d\Gamma_D}{dw} (B\dec D\ell\nu)= \frac{G_F^2|
V_{cb}|^2}{48\pi^3}{\cal K_D}(w){\cal G}(w)^2,
\end{equation}
where $w$ is the inner product of the B and D meson 4-velocities,
${\cal K_D}(w)$ is the phase space and the form factor ${\cal
G}(w)$ is generally expressed as the product of a normalization
factor ${\cal G}(1)$ and a function, $g_D(w)$, constrained by
dispersion relations \cite{grinstein}.

The strategy to extract ${\cal G}(1)|V_{cb}|$ is identical to that
used for the $B\dec D^{\star} \ell \nu$ decay. However both theory
and experiments have additional difficulties in dealing with this
channel. From the theoretical standpoint, the non-perturbative
expansion includes  $1/m_b$ and $1/m_c$ terms, as there is no
suppression mechanism. Moreover, this is a decay that is
experimentally challenging as it is difficult to isolate from the
larger $D^{\star} \ell \bar{\nu}$ final state.

Belle \cite{belle-dplnu} and ALEPH \cite{ALEPH_vcb} studied the
$\bar{B}^0\dec D^+ \ell^- \bar{\nu}$ channel, while CLEO
\cite{cleo-dplnu} studied both $B^+\dec D^0 \ell^+ \bar{\nu}$ and
$ \bar{B^0}\dec D^+ \ell^- \bar{\nu}$ decays. The results scaled
to common inputs are shown in Table~\ref{t:D}.  Averaging
\cite{lepvcb} the data in Table~\ref{t:D}, using the procedure  of
\cite{lepvcb}, we get ${\cal G}(1) |V_{cb}| = (41.3  \pm
4.0)\times 10^{-3}$ and $\rho_D^2 = 1.19 \pm 0.19,$ where
$\rho_D^2$ is the slope of the form-factor  at zero recoil given
in \cite{CLN}.

\begin{table}
\begin{tabular}{lcc} \hline
{Experiment}  & {${\cal G}(1)|V_{cb}|(\times 10^{3})$} &
{$\rho^2_D$} \\
\hline
 ALEPH       &  40.0$\pm$  10.0$\pm$ 6.4 & 1.02 $\pm$ 0.98$\pm$ 0.37  \\
 Belle       &  41.8$\pm$  4.4$\pm$ 5.2 & 1.12$\pm$ 0.22$\pm$ 0.14 \\
 CLEO        &  44.9$\pm$  5.8$\pm$ 3.5 & 1.27$\pm$ 0.25$\pm$ 0.14 \\
\hline
World average &  42.1 $\pm$ 3.7 & 1.15$\pm$ 0.16 \\
\hline
\end{tabular}
\caption{Experimental results corrected to common inputs b and
world average \cite{hfag}. $\rho_D^2$ is the slope of the
form-factor given in \cite{CLN} at zero recoil. \label{t:D}}
\end{table}

The theoretical predictions for ${\cal G}(1)$ are consistent: a
quark model evaluation gives $1.03 \pm
0.07$\cite{ScoraIsgurPRD5295}, and a recent heavy quark sum rule
calculation \cite{sumr} gives $1.04 \pm 0.02 \pm \delta_{exp}$,
where $\delta_{exp}$ represents the error in $\mu_{\pi}^2(1\rm
GeV)$, defined in the next section.
A 
quenched lattice calculation gives ${\cal G}(1)=1.058
^{+0.021}_{-0.017}$\cite{JLQCDPRL8299}, where the errors do not
include the uncertainties induced by the quenching approximation
and lattice spacing. Using ${\cal G}(1)=1.04\pm\ 0.07$, we get
$|V_{cb}| = (40.5 \pm 3.6_{\rm exp} \pm 2.7_{\rm theo}) \times
10^{-3},$ consistent with the value extracted from $ B\to
D^{\star} \ell \nu$ decay, but with a larger uncertainty.

\section{$|V_{cb}|$ from the inclusive decay $B\rightarrow X_c \ell \bar{\nu}$.}

  The decay $B\rightarrow
X_c \ell \bar{\nu}$ is an alternative experimental approach to
extract $|V_{cb}|$. In this case, the Operator Product Expansion
(OPE) is the theoretical tool used. It yields the heavy quark
inclusive decay rates as an asymptotic series in inverse powers of
the heavy quark mass. More precisely, several mass scales are
relevant: the b quark mass $m_b$, the c quark mass $m_c$ and the
energy release $E_r\equiv m_b -m_c$ \cite{kolya-2001}. The
uncertainties in the predicted $\Gamma _{sl}/|V_{cb}|$ have been
discussed in numerous theoretical papers
\cite{benson-et-al-durham03}. However, the theory needs to provide
predictions on independent observables that can be used to
validate its accuracy. Experimental input include the semileptonic
width, as well as the determination of the theoretical parameters
governing the hadronic matrix element, discussed below.

 The key parameter in the theoretical expression for the
semileptonic width is $m_b$. As the bare quark mass is affected by
perturbative and non-perturbative contributions, considerable
attention has been devoted to its proper definition \cite{luke02},
\cite{hoang}. Similarly, $m_c$ is a parameter in the hadronic
matrix element and, recently, it has been argued
\cite{benson-et-al-durham03} that extracting it from the
relationship between $(m_b-m_c)$ and the spin averaged meson mass
difference $(\bar{M_B}- \bar{M_D})$ \cite{falk} may be inadequate.

 The leading
non-perturbative corrections arise only to order $1/m_b^2$ and are
parameterized by the quantities $\mu_{\pi}^2\ ({\rm or}\ -\lambda
_1)$ \cite{falk}, \cite{kolya} related to the expectation value of
the kinetic energy of the $b$ quark inside the $b$ hadron, and
$\mu _G^2\ ({\rm or}\ \lambda _2)$ \cite{falk}, \cite{kolya}
related to the expectation value of the chromomagnetic operator.
Quark-hadron duality is an important {\it ab initio} assumption in
these calculations. While several authors \cite{bigiduality} argue
that this ansatz does not introduce appreciable errors as they
expect that duality violations affect the semileptonic width only
in high powers of the non-perturbative expansion, other authors
recognize that an unknown correction may be associated with this
assumption \cite{buchalla}. Arguments supporting a possible
sizeable source of errors related to the assumption of
quark-hadron duality have been proposed \cite{nathan}.

I will start the discussion with the experimental studies of the
moments of inclusive distributions. Most of the experimental
studies have focused on the lepton energy and the invariant mass
$M_X$ of the hadronic system recoiling against the
lepton-$\bar{\nu}$ pair. CLEO published the first measurement of
the moments of the $M_X^2$ distributions. This analysis includes a
1.5 GeV/c lepton momentum cut, that allows them to single out the
desired $b \rightarrow c \ell^- \bar{\nu}$ signal from the
``cascade" $b\rightarrow c \rightarrow s \ell ^+\nu$ background
process. The hermeticity of the CLEO detector is exploited to
reconstruct the $\nu$ 4-momentum vector. Moreover, the $B\bar{B}$
pair is produced nearly at rest and thus it allows a determination
of $M_X$ from the $\nu$ and $\ell$ momenta. They obtain
$<M_X^2-\bar{M_D}^2> = 0.251 \pm 0.066\ {\rm GeV}^2$ and
$<(M_X^2-\bar{M_D}^2>)^2> = 0.576 \pm 0.170 \ {\rm GeV}^4$, where
$\bar{M_D}$ is the spin-averaged mass of the $D$ and $D^{\star}$
mesons. The lepton momentum cut may reduce the accuracy of the OPE
predictions, because restricting the kinematic domain may increase
quark-hadron duality violations. The shape of the lepton spectrum
provides further constraints on OPE. Moments of the lepton
momentum with a cut $p_{\ell}^{CM}\ge 1.5$ GeV/c have been
measured by the CLEO collaboration \cite{prd2003}. The two
approaches give consistent results, although the technique used to
extract the OPE parameters has still relatively large
uncertainties associated with the $1/m_b^3$ form factors. The
sensitivity to $1/m_b^3$ corrections depends upon which moments
are considered. Bauer and Trott \cite{bt-2001} have performed an
extensive study of the sensitivity of lepton energy moments to
non-perturbative effects. In particular, they have proposed
``duality moments," very insensitive to neglected higher order
terms. The comparison between the CLEO measurement of these
moments\cite{prd2003} and the predicted values shows a very
impressive agreement:
\begin{eqnarray}
D_3&\equiv& {\int_{1.6\,{\rm GeV}} E_\ell^{0.7}{d\Gamma\over
dE_\ell}\,dE_\ell\over \int_{1.5\,{\rm GeV}}
E_\ell^{1.5}{d\Gamma\over
dE_\ell}\,dE_\ell}=
\cases{0.5190\pm 0.0007&(T)\cr
0.5193\pm 0.0008&(E)}
\nonumber\\
D_4&\equiv& {\int_{1.6\,{\rm GeV}} E_\ell^{2.3}{d\Gamma\over
dE_\ell}\,dE_\ell\over \int_{1.5\,{\rm GeV}}
E_\ell^{2.9}{d\Gamma\over dE_\ell}\,dE_\ell}=\cases{0.6034\pm
0.0008&(T)\cr
0.6036\pm 0.0006&(E)} \nonumber\\
\end{eqnarray}
(where ``T" and ``E" denote theory and experiment, respectively).

More recently, both CLEO and BaBar explored the moments of the
hadronic mass $M_X^2$ with lower lepton momentum cuts. In order to
identify the desired semileptonic decay from background processes
including cascade decays, continuum leptons and fake leptons, CLEO
performs a fit for the contributions of signal and backgrounds to
the full three-dimensional differential decay rate distribution as
a function of the reconstructed quantities $q^2$, $M_X^2$,
$\cos{\theta _{W\ell}}$. The signal includes the components
$B\rightarrow D \ell \bar{\nu}$, $B\rightarrow D^{\star} \ell
\bar{\nu}$, $B\rightarrow D^{\star\star} \ell \bar{\nu}$,
$B\rightarrow X_c \ell \bar{\nu}$ non-resonant and $B\rightarrow
X_u \ell \bar{\nu}$. The backgrounds considered are: secondary
leptons, continuum leptons and fake leptons. BaBar uses a sample
where the hadronic decay of one $B$ is fully reconstructed and the
charged lepton from the other $B$ is identified. In this case the
main sources of systematic errors are the uncertainties related to
the detector modeling and reconstruction.

\begin{figure}[htb]
\epsfig{file=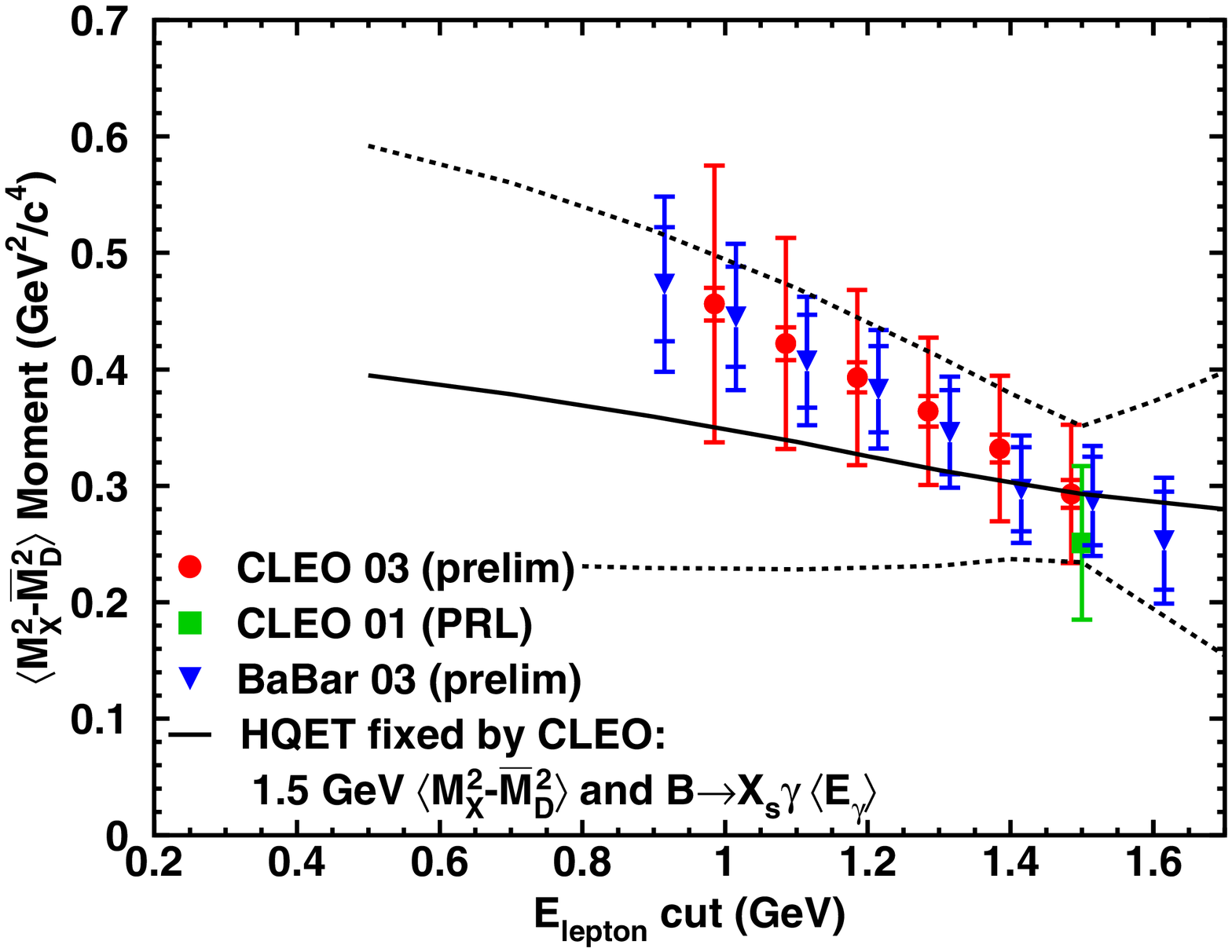,width=150mm} \caption{The results of
the recent CLEO analysis \cite{newmxmom} compared to previous
measurements \cite{ref:oldhadmom,ref:babarmoms} and the HQET
prediction. The theory bands shown in the figure reflect the
variation of the experimental errors on the two constraints, the
variation of the third-order HQET parameters by the scale $(0.5\
{\rm GeV})^3$, and variation of the size of the higher order QCD
radiative corrections \cite{bauer-et-al-2002}.}
\label{mom-leptons}
\end{figure}

Fig.~\ref{mom-leptons} shows the extracted $<M_X^2-\bar{M}_D^2>$
moments as a function of the minimum lepton momentum cut from
these two measurements, as well as the original measurement with
$p_{\ell}\ge 1.5$ GeV/c. The results are compared with theory
bands that reflect experimental errors, $1/m_b^3$ correction
uncertainties and uncertainties in the higher order QCD radiative
corrections \cite{bauer-et-al-2002}. The CLEO and BaBar results
are consistent and show an improved agreement with theoretical
predictions with respect to earlier preliminary results
\cite{babar-ichep02}. Moments of the $M_X$ distribution without an
explicit lepton momentum cut have been extracted from preliminary
DELPHI data \cite{battaglia} and give consistent results.

\vspace{1cm}

\begin{figure}[htb]
\epsfig{file=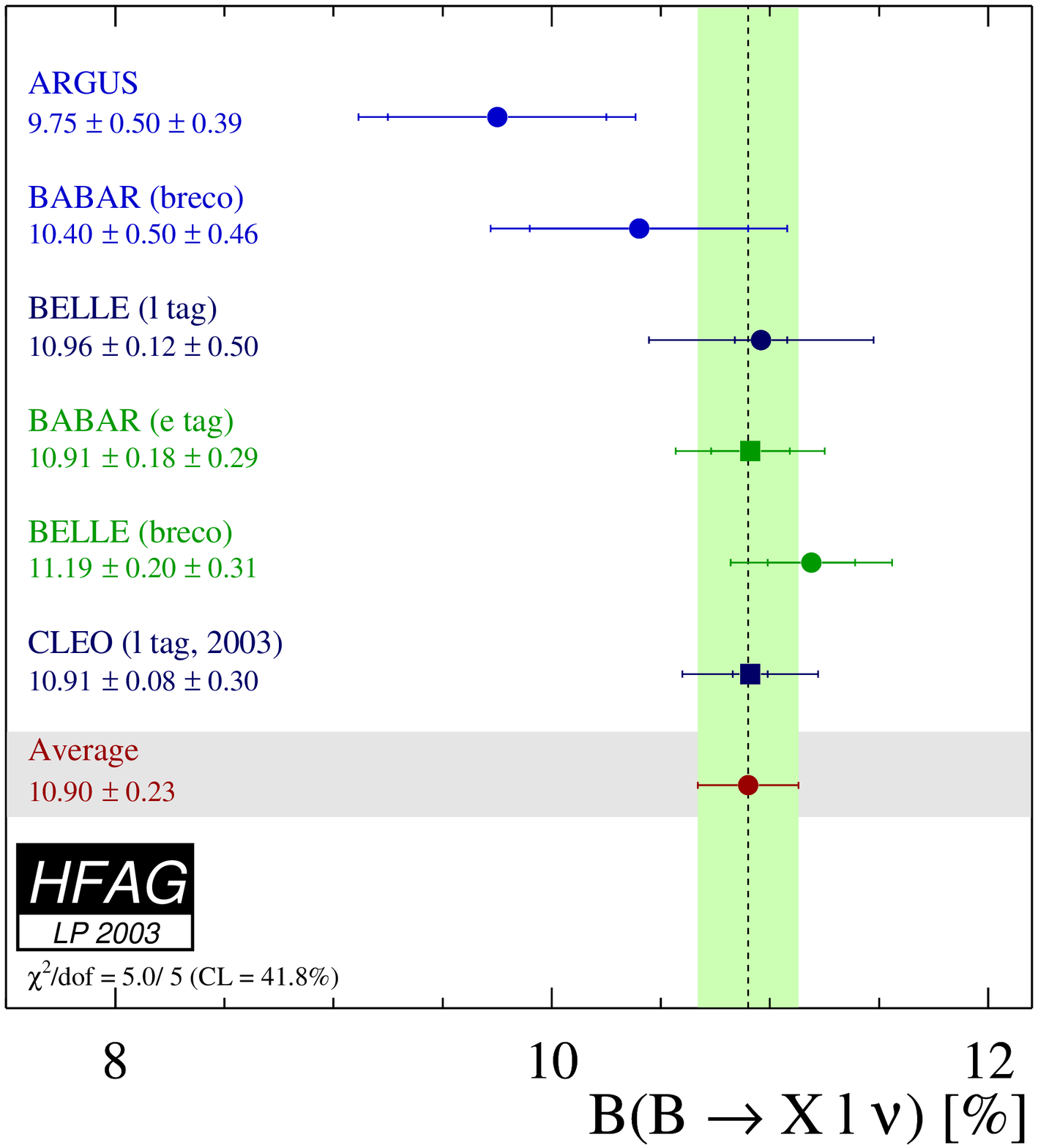,width=150mm} \caption{Summary of
model independent semileptonic branching fraction measurements
performed at the $\Upsilon(4S)$ center-of-mass energy.}
\label{sl-inclusive-4s}
\end{figure}

The second element needed to extracted $|V_{cb}|$ with this method
is the measured semileptonic width. Experiments operating at the
\ups\ center-of-mass energy use a dilepton sample to separate the
decay process $b\rightarrow c \ell^- \bar{\nu}$ (primary leptons)
from the $b\rightarrow c\rightarrow s \ell ^+\nu$ (cascade
leptons). This technique allows a direct determination of the
primary lepton spectrum over almost all the kinematically allowed
range. Thus, the semileptonic branching fraction extracted from
this measurement has almost no model dependence.
Fig.~\ref{sl-inclusive-4s} shows a summary of the \ups\
measurements of inclusive semileptonic branching fractions. The
overall experimental error is of the order of 2\%. Different
extractions of the HQE non-perturbative parameters cannot be
combined in a straighforward manner because they use different
methods to estimate the theoretical uncertainties and they do not
fully agree \cite{ref:babarmoms}. I will choose a representative
set of parameters \cite{prd2003} and obtain:
\begin{equation} |{\mbox{V}}_{cb}|=(41.5\pm 0.4|_{\Gamma_{sl}}\pm
0.4|_{\lambda_1, \bar{\Lambda}}\pm 0.9|_{th})\times 10^{-3},
\end{equation}
where the first uncertainty is from the experimental value of the
semileptonic width, the second uncertainty is from the HQE
parameters (${\lambda}_{1}$ and $\bar{\Lambda}$) and the third
error is the theoretical uncertainty in the hadronic matrix
element. No quantitative account is given for possible quark
hadron duality violation. The present difference between the two
values of $|V_{cb}|$ obtained from $B\dec D^{\star} \ell
\bar{\nu}$ and from inclusive semileptonic branching fraction
measurements may be used to make a very rough estimate of the non
quantified errors in the inclusive determination of $|V_{cb}|$ at
a level of about 6\%.

\section{Conclusions}

The values of $\vcb$ obtained both from the inclusive and
exclusive method agree within errors. The value of $\vcb$ obtained
from the analysis of the $B\rightarrow D^{\star}\ell \nu$ decay
is:
\begin{equation}
|V_{cb}|_{exclusive} = (40.2 \pm 0.9_{exp} \pm 1.8_{theo}) \times
10^{-3}
\end{equation}
where the first error is experimental and the second error is from
the $1/m_b^2$ corrections to ${\cal F}(1)$. The value of $\vcb$,
obtained from inclusive semileptonic branching fractions is:
\begin{equation}
|V_{cb}|_{incl} = (41.4 \pm 0.5_{exp} \pm
0.4_{\lambda_1,\overline{\Lambda}}\pm 0.9_{theo}) \times 10^{-3},
\end{equation}
where the first error is experimental, the second error is from
the measured values of $\lambda_1$, and $\overline{\Lambda}$,
assumed to be universal up to higher orders, and the last from
$1/m_b^3$ corrections and $\alpha_s$. Non-quantified uncertainties
are associated with a possible quark-hadron duality violation. An
estimate through a comparison between these two results implies an
additional uncertainty of the order of 6\%. For this reason I
choose not to average these two numbers, but quote a conservative
estimate based on $|V_{cb}|$ exclusive.

High precision tests of HQET and more precise assessment of
quark-hadron duality in inclusive semileptonic decays are needed
to achieve the ultimate accuracy in this measurement.

\section{Acknowledgments}
I would like to thank the organizers for a very interesting and
lively conference. I would also like to thank C. Boulahouache and
S. Stone for interesting discussions. This work was supported from
the US National Science Foundation.

\end{document}